\documentclass[preprint,aps,eqsecnum,11p,nofootinbib]{revtex4}
\usepackage{epsfig}

\newcommand{\be}{\begin{equation}}
\newcommand{\ee}{\end{equation}}
\newcommand{\bea}{\begin{eqnarray}}
\newcommand{\eea}{\end{eqnarray}}

\begin{document}
\title{Evolution of primordial perturbations and a fluctuating decay rate.}
\author{$^{a}$~Anupam Mazumdar, and  $^{b}$~Marieke Postma}

\affiliation{$^{a}$ CHEP, McGill University, Montr\'eal, QC, H3A 2T8, Canada\\
$^{b}$ The Abdus Salam International Centre for Theoretical Physics, 
Strada Costiera, I-10, 34100 Trieste, Italy}
\begin{abstract}
We present a gauge invariant formalism to study the evolution of
curvature perturbations during decay of the inflaton or some other
field that dominates the energy density.  We specialize to the case
where the total curvature perturbation arises predominantly from a
spatially fluctuating decay rate.  Gaussian fluctuations in the decay
rate are present if the inflaton coupling or mass depends on some
light field, which fluctuates freely during inflation. The
isocurvature fluctuations of the light field seeds the total curvature
perturbations, whose amplitude freezes after inflaton decay. We also
analyze the resulting curvature perturbation starting with
non-negligible values of the initial inflaton curvature.  Lastly, we
study the effects of the energy density stored in the light field.
\end{abstract}
\maketitle

\section{Introduction}

Inflation is the main contender for an explanation of the observed
adiabatic density perturbations with a nearly scale invariant
spectrum~\cite{Linde}. However, recently, alternative mechanisms for
generating the density perturbations have also been discussed. In the
curvaton scenario, isocurvature perturbations of some light
``curvaton'' field are converted into adiabatic perturbations in the
post-inflationary universe~\cite{curvaton}.  

Another interesting proposal is that the perturbations could be
generated from the fluctuations of the inflaton coupling to the
Standard Model degrees of freedom~\cite{Dvali,Kofman,epm}. It has been
argued that the coupling strength of the inflaton to ordinary matter
or the inflaton mass, instead of being a constant, could depend on the
vacuum expectation values (VEV) of the various fields in the
theory. If these fields are light during inflation their quantum
fluctuations will lead to spatial fluctuations in the inflaton decay
rate. As a consequence, when the inflaton decays, adiabatic density
perturbations will be created because fluctuations in the decay rate
translate into fluctuations in the reheating temperature. A variation
of the above scenario is that the inflaton decays into heavy particle,
whose decay is governed by a varying decay rate.  If these particles
come (close) to dominating the energy density of the universe,
curvature perturbations are produced~\cite{Dvali2}.  The light fields
whose fluctuations leads to a fluctuating decay rate could be normal
matter fields, such as the flat directions in the minimal
supersymmetric Standard Model (MSSM), or a right handed
neutrino~\cite{Mazumdar}.
            
In this paper we will discuss the evolution of the curvature
perturbation during decay, in the case of a fluctuating decay rate.
We will use the gauge-invariant formalism for metric perturbations as
developed by Bardeen~\cite{Bardeen}, for a review see 
Ref.~\cite{Brandenberger}. The curvature perturbation on
large scales (defined on a uniform density spatial hypersurface)
$\zeta$ remains constant for purely adiabatic perturbations. But
$\zeta$ can change on large scales due to a non-adiabatic pressure
perturbation.  In a multi-fluid system such pressure perturbations can
be important in the presence of relative entropy perturbations.  Said
in another way, entropy (or isocurvature) perturbations can feed the
curvature perturbation.  Therefore, to compute the final perturbation
spectrum in a multi fluid system it is necessary to keep track of the
evolution and perturbations of various fluids.

In the case of a fluctuating decay rate, the multi fluid system
consist of the decaying field (the inflaton, or some other field which
has dominant energy density), the radiation bath, and the light or
flat direction field that is responsible for the fluctuating decay
rate. The relative curvature perturbation between the inflaton and the
flat direction acts as a source for the total curvature perturbations.
This can be understood intuitively from the fact that fluctuations in
the inflaton decay rate leads to fluctuations in the reheat
temperature of the universe, given by 
$T_{\rm rh}\sim \lambda\sqrt{m_{\phi}M_{p}}$, where $m_{\phi}$ is 
the mass of the inflaton
and $M_{P} =2.436\times 10^{18}$~GeV is the reduced Planck mass. The
fluctuations in $\Gamma$ can be translated into fluctuations in the
energy density of a thermal bath with 
$\delta \rho_{\gamma}/\rho_{\gamma}=-(2/3)\delta\Gamma/\Gamma$~\cite{Dvali}. 
The factor $2/3$ appears due to red-shift of the modes during the decay of the
inflaton whose energy still dominates.

The inflaton decay rate is $\Gamma\sim m_\phi\lambda^2$. The decay 
rate fluctuates if either $\lambda$ or $m_\phi$
is a function of a fluctuating light field. The former case can
e.g. arise in the MSSM through terms in the superpotential of the form
\begin{equation}
\label{superpot}
W \ni \lambda_h \phi \bar{H} H +
\phi \frac{q}{M} q q + \phi \frac{q_c}{M} q_c q_c +
\phi \frac{h}{M} q q_c,
\end{equation}
with the inflaton field $\phi$ a standard model singlet, $H$ and
$\bar{H}$ the two Higgs doublets, and $q$ and $q_c$ the quark and
lepton superfields and their anti-particles. $M$ is some cut-off scale
which could be the GUT scale or the Planck scale. In this paper we
will not be concerned with the particulars of the decay products and
the light fields.  What we will take from the above example though, is
that there can be two kinds of couplings:
\be
\label{rate}
\lambda = \left \{ 
\begin{array}{lll} 
&\lambda_{0} \left( 1 +\frac{S}{M}+...\right), 
& \qquad {\rm direct~decay}.
\\
&\lambda_{0} \left( \frac{S}{M} + ... \right), 
& \qquad {\rm indirect~decay},
\end{array}
\right.
\ee
with the ellipses standing for higher order terms.  In the MSSM
example, the ``direct'' decay mode corresponds to inflaton decay into
Higgs fields, whereas the ``indirect'' decay mode corresponds to decay
into (s)quarks and anti-(s)quarks. $S$ is the expectation value of a
light fluctuating field. Effective couplings of this form can result
from integrating out heavy particles.  Another possibility is that the
inflaton, or some other field that dominates the energy density of the
universe (e.g. non-relativistic inflaton decay products), has an
effective mass generated through a coupling to the flat direction:
$m_\psi = \lambda S.$ 
The fluctuation in the decay rate for the various cases is
\be
\label{decay}
\frac{\delta \Gamma}{\Gamma} = \left \{ 
\begin{array}{lll} 
& 2\frac{\delta S}{M}, 
& \qquad {\rm direct~decay}.
\\
& 2 \frac{\delta S}{S}, 
& \qquad {\rm indirect~decay}.
\\
& \frac{\delta S}{S},
& \qquad {\rm fluctuating~mass}.
\end{array}
\right.  \ee

\section{Background equations}

The background equations of motion governing the dynamics of the
inflaton, the radiation bath, and the flat direction field in 
a Robertson Walker universe are
\bea
\label{dyn1}
\dot{\rho}_\phi &=& - \rho_\phi (3H + \Gamma)\,, \\
\label{dyn2}
\dot{\rho}_\gamma &=& - 4 H \rho_\gamma + \Gamma \rho_\phi\,, \\
\label{dyn3}
\dot{\rho}_S &=& -3 H \rho_S (1+\omega)\,. 
\eea
where a dot denotes differentiation w.r.t. coordinate time. The Hubble
parameter $H=\dot a/a$, with $a$ the scale factor of the universe, is
given by
\be H^2 = \frac{1}{3M_{P}^2} (\rho_\phi + \rho_\gamma + \rho_S).  \ee
Here $\omega = p_S /\rho_S$ is the equation of state parameter for the
flat direction field $S$.  When $H \gg m_S$, the flat direction field
is over damped and remains effectively frozen, and $\omega \to -1$.
For $H \lesssim m_S$, the flat direction field oscillates in its
potential. For a quadratic potential $\omega = 0$ averaged over one
oscillation. To determine $\omega$ and the decay width $\Gamma(S)$ we
need the evolution of $S$:
\be
\ddot{S} + 3 H \dot{S} + V_{,S} = 0
\ee
with $V_{,S} = \partial V / \partial S$. Then 
\be
\omega = \frac{p_S }{ \rho_S} = \frac{ {\dot{S}^2}/{2} - V(S)}
{ {\dot{S}^2}/{2} +V(S)}.
\ee

It is useful to introduce the following dimensionless parameters
\be
\Omega_\alpha = \frac{\rho_\alpha}{\rho},\qquad
s = \frac{S}{m},\qquad
g = \frac{\Gamma}{m_\phi}, \qquad
h = \frac{H}{m_\phi}. 
\ee
We leave the mass scale $m$ unspecified for now, it can be taken
$m_S$.  Note however that if finite temperature effects or effects of
non-renormalizable operators are taken into account, then $m_S$ is
time-dependent and we cannot set $m = (m_S)_{\rm eff}$.  Further, we
eliminate $\Omega_\gamma$ from all equations by use of the Friedman
constraint:
\be 
\Omega_S + \Omega_\phi + \Omega_\gamma = 1 
\ee
The dimensionless background equations are then 
\bea
\label{dim1}
\Omega'_\phi &=& \Omega_\phi 
\Big( 1- \frac{g}{h} - \Omega_\phi + \Omega_S (3\omega -1) \Big) \\
\label{dim2}
\Omega'_S &=&  \Omega_S \Big( (1 -3 \omega) - \Omega_\phi + 
(3 \omega -1) \Omega_S \Big) \\
\label{dim3}
h' &=& \frac{h}{2} \Big( -4+\Omega_\phi + (1-3\omega) \Omega_S \Big) \,.
\eea
and
\be
s'' + \left( \frac{h'}{h}+3 \right) s' + \left(\frac{V_{,s}(s)}{h^2}\right) s =0.
\ee
Here, prime denotes differentiation with respect to the number of
e-foldings $N \equiv \ln a$.


\section{Gauge invariant perturbations}

We are interested in the long wavelength regime of the perturbations
where the comoving scale is much larger than the Hubble horizon.
Following Ref.~\cite{Bardeen} we define the curvature perturbation
$\zeta$ on spatial slices of uniform density $\rho$ with the line
element~\footnote
{Linear perturbations around a flat FRW metric can be defined by the
line element $ds^2 = -(1+2\phi) dt^2 + 2aB_{,i} dtdx^{i} + a^2 \left[
(1-2\psi) \delta_{ij} + 2E_{,ij} \right] dx^{i}dx^{j}$, where we have
used the notation of Ref.~\cite{Brandenberger} for the gauge dependent
curvature perturbation $\psi$, the lapse function $\phi$, and the
scalar shear $\chi=a^2\dot E-aB$.  The quantity $\zeta$ is related
to the curvature perturbation $\psi$, on a generic slicing, 
by $\zeta = -\psi - H \delta\rho / \dot\rho$.}
\begin{equation}
ds^2= a^2(t)\left(1+2\zeta \right)\delta_{ij}dx^{i}dx^{j}\,,
\end{equation}
where $a$ is the scale factor. The time evolution of the curvature 
perturbation on large scales is~\cite{Kodama,Bellido,Wands2}
\begin{equation}
\label{curv0}
\dot\zeta=-\frac{H}{\rho+P}\delta P_{{\rm nad}}\,,
\end{equation}
where $P_{{\rm nad}}\equiv \delta P-c_{s}^2\delta\rho$ is the non-adiabatic
pressure perturbation. The adiabatic sound speed is 
$c_{s}^2=\dot P/\dot\rho$, where $P$ and $\rho$ are the total pressure
and energy density respectively. For a single field 
$\delta P_{{\rm nad}}=0$, and therefore on large scales the 
curvature perturbation is pure adiabatic in nature 
with $\zeta={\rm constant}$.

The total curvature perturbation Eq.~(\ref{curv0}) can be written in 
terms of the various components
\begin{equation}
\label{curv1}
\zeta=\sum_{\alpha}\frac{\dot\rho_{\alpha}}{\dot\rho}\zeta_{\alpha}\,.
\end{equation}
In our case $\alpha=\phi,S,\gamma$. Isocurvature or entropy
perturbations describe the difference between the curvature
perturbations \cite{Kodama} 
\be
\label{iso}
S_{\alpha \beta} =  3\left(\zeta_{\alpha}-\zeta_{\beta}\right)=-3 H \left( 
\frac{\delta \rho_\alpha}{\dot{\rho}_\alpha}
-\frac{\delta \rho_\beta}{\dot{\rho}_\beta}
\right)\,.
\ee
With the help of Eqs.~(\ref{curv1},\ref{iso}), we can obtain a 
useful relationship
\begin{equation}
\zeta_{\alpha}=\zeta+\frac{1}{3}\sum_{\beta}\frac{\dot\rho_{\beta}}{\dot\rho}
S_{\alpha\beta}\,,
\end{equation}
where $\alpha,\beta=\phi,S,\gamma$.

In the presence of more than one field the non-adiabatic pressure
perturbation can be non-zero $\delta P_{{\rm nad}}\neq 0$, and
therefore the total curvature perturbations evolve in time. The
non-adiabatic pressure has two contributions, one is from the
intrinsic pressure perturbations and the other from the relative
pressure perturbations~\cite{Kodama,Wands2}.
\begin{equation}
\delta P_{{\rm nad}}\equiv 
\sum_{\alpha}\delta P_{{\rm intr}, \alpha}+\delta P_{\rm rel}\,,
\end{equation}
where $\delta P_{{\rm intr},\alpha}\equiv\delta
P_{\alpha}-c^2_{\alpha}\delta\rho_{\alpha}$, and
$c^2_{\alpha}\equiv\dot P_{\alpha}/\dot\rho_{\alpha}$ is the adiabatic
speed of sound for a respective fluid. Note that the intrinsic
pressure perturbation vanishes for a fixed equation of state. Thus,
$\delta P_{{\rm intr},\gamma}=0$, and (during the stage of inflaton
oscillations) $\delta P_{{\rm intr},\phi}=0$. For a fixed $\omega$
parameter, also the intrinsic pressure perturbation for the flat
direction vanishes $\delta P_{{\rm intr}, S} =0$.  However the
relative pressure perturbation is non-zero and can be expressed in
terms of the various entropy perturbations as~\cite{Kodama}
\begin{equation}
\label{rel}
\delta P_{{\rm rel}}\equiv -\frac{1}{6H\dot\rho}\sum_{\alpha.\beta}\dot\rho_{\alpha}
\dot\rho_{\beta}\left(c_{\alpha}^2-c^2_{\beta}\right)S_{\alpha\beta}\,.
\end{equation}
Using Eqs.~(\ref{curv0},\ref{curv1},\ref{iso},\ref{rel}), the
evolution of the total curvature perturbation becomes
\be
\label{zeta0}
\dot{\zeta} 
 =  \frac{H}{\dot{\rho}^2} \left( \frac{1}{3} \dot{\rho}_\phi  
\dot{\rho}_\gamma
S_{\phi \gamma} - (\omega - \frac{1}{3}) \dot{\rho}_S  \dot{\rho}_\gamma
S_{S \gamma} + \omega  \dot{\rho}_\phi  \dot{\rho}_S S_{\phi S} \right)\,. 
\ee

During the decay of the inflaton in relativistic species there is
non-adiabatic energy transfer from the inflaton to the radiation bath,
see Eqs.~(\ref{dyn1},\ref{dyn2}):
\begin{eqnarray}
\label{qnad}
Q_{\phi} &=&-\Gamma\rho_{\phi}\,, \\ 
Q_{\gamma}&=&\Gamma\rho_{\phi}\,.
\end{eqnarray}
Due to fluctuations in $\Gamma$, the non-adiabatic energy transfer is
also subject to perturbations. Therefore, the evolution of the
individual curvature perturbations $\zeta_{\phi}$ and $\zeta_{\gamma}$
is determined not only by non-adiabatic pressure perturbations but
also by the non-adiabatic energy transfer perturbations $\delta
Q_{{\rm nad},\alpha}$. Analogous to the non-adiabatic pressure
perturbations, these can be split in an intrinsic and relative
part~\cite{Kodama,Wands2}
\begin{equation}
\delta Q_{{\rm nad},\alpha}\equiv \delta Q_{{\rm intr},\alpha}+\delta Q_{{\rm rel},\alpha}\,.
\end{equation}
The intrinsic non-adiabatic energy transfer is defined as
\begin{equation} 
\label{intr}
\delta Q_{{\rm intr},\alpha}\equiv \delta Q_{\alpha}-\frac{\dot Q_{\alpha}}{
\dot\rho_{\alpha}}\delta\rho_{\alpha}\,,
\end{equation}
whose contribution is zero if $Q_{\alpha}$ is a function of local
energy transfer $\delta Q_{\alpha}=\dot
Q_{\alpha}\delta\rho_{\alpha}/\dot\rho_{\alpha}$
\cite{Kodama,Wands2}. The non-zero contributions are
\bea
\delta Q_{{\rm int},\phi} 
&=& -\frac{ \dot{\Gamma} }{3H} \rho_\phi S_{\phi S}\,, \\
\delta Q_{{\rm int},\gamma} 
&=& \frac{ \dot{\Gamma}}{3H} \rho_\phi S_{\gamma S}
- \frac{\Gamma}{3H} \dot{\rho}_\phi  S_{\phi \gamma}\,,
\eea
where we have used $\delta\Gamma/\dot{\Gamma}=\delta{S}/\dot{S}$, see
Eq.~(\ref{rate}), to write the expressions in terms of $S_{\alpha S}$.

The relative non-adiabatic energy transfer is given by~\cite{Kodama,Wands2}
\begin{equation}
\label{relative}
\delta Q_{{\rm rel},\alpha}\equiv
-\frac{Q_{\alpha}}{6H\rho}\sum_\beta\dot\rho_{\beta}
S_{\alpha\beta}\,.
\end{equation} 
The non-zero components are 
\bea
\label{relative1}
\delta Q_{{\rm rel},\phi} &=& \frac{\Gamma \rho_\phi}{6 H \rho} 
\left( \dot{\rho}_S S_{\phi S} + \dot{\rho}_\gamma S_{\phi \gamma}\right)\,, \\
\label{relative2}
\delta Q_{{\rm rel},\gamma} &=& - \frac{\Gamma \rho_\phi}{6 H \rho} 
\left( \dot{\rho}_\phi S_{\gamma \phi} +
\dot{\rho}_S S_{\gamma S} \right)\,.
\eea
The evolution of the individual gauge invariant curvature
perturbations (on uniform density hypersurfaces) is fed by the
non-adiabatic perturbations through ~\cite{Kodama,Wands2}
\begin{equation}
\label{acurv}
\dot\zeta_{\alpha}=\frac{3H^2\delta P_{{\rm intr},\alpha}}{\dot\rho_{\alpha}}-
\frac{H\delta Q_{{\rm nad},\alpha}}{\dot\rho_{\alpha}}\,.
\end{equation}
The evolution equations of the curvature perturbations then are 
\bea
\label{curvature1}
\dot{\zeta}_\phi &=& 
\frac{\rho_\phi} {3 \dot{\rho}_\phi } 
\left( \dot{\Gamma} - \frac{\Gamma \dot{\rho}_S} {2 \rho} \right) S_{\phi S} - 
\frac{\Gamma \rho_\phi \dot{\rho}_\gamma}
{6 \rho \dot{\rho}_\phi} S_{\phi \gamma}\,, \\
\label{curvature2}
\dot{\zeta}_\gamma &=& 
\frac{\rho_\phi} {3 \dot{\rho}_\gamma } 
\left( -\dot{\Gamma} + \frac{\Gamma \dot{\rho}_S} {2 \rho}\right) 
S_{\gamma S} +\frac{\Gamma \dot{\rho}_\phi}
{3 \dot{\rho}_\gamma} \left( 1 - \frac{\rho_\phi} {2 \rho} \right) 
S_{\phi \gamma}\,, \\ 
\label{curvature3}
\dot{\zeta}_S &=& \frac{3 H^2 \delta P_{{\rm intr},S}}{\dot{\rho}_{S}}\,.
\eea

The equation for $\dot{\zeta}_S$ needs some explanation. First of all,
note that the light field $S$ does not undergo energy transfer, unlike
the inflaton and the radiation bath.  If the equation of state
parameter is a constant, as it is for example during $S$ oscillations,
the non-adiabatic pressure perturbation vanishes and $\zeta_S$ remains
constant. This just reflects the fact that $\delta S/S
\propto\delta\rho_S/\rho_S$ is constant for a quadratic potential, as
$S$ and $\delta S$ follow the same equation of motion.

During slow roll the equation of state parameter $\omega$ is changing
with time and the non-adiabatic pressure perturbation is non-zero.  If
the flat direction field is rolling in a potential dominated by
non-renormalizable terms, this can lead to appreciable damping of
$(\delta S / S)$, as discussed in Ref.~\cite{epm}. Therefore, we will
assume a quadratic potential for $S$ with non-renormalizable terms
negligible small.  This assumption is justified if the VEV of $S$ is
small enough, $\frac{1}{2} m_{S}^2 S^2 \ll \lambda^2
S^{2n-2}/M^{2n-6}$, with $n \ge 4$.  Small VeVs are naturally
obtained in low scale inflation.

However, this still prohibits us from taking the `frozen' limit
$\dot{S} \to 0$ in a naive way.  The problem is that in this limit
$\omega \to -1$, and $\zeta_S \to \infty$ becomes ill defined. To do
things properly one has to take the slow roll into account.  We write
$\omega = -1 + \epsilon$, with $\epsilon=2T/V$ in the slow roll limit
($\ddot{S} \to 0$, and $T \ll V$). Here $T = \frac{1}{2}\dot{S}^2$ the
kinetic energy, and $V = \frac{1}{2} m^2 S^2$ is the potential
energy. Then
\be
P_{{\rm intr},S} = - \frac{\dot{\omega} \rho_S}{\dot{\rho}_S} \delta \rho_S
= \epsilon \rho_S \frac{\delta \rho_S}{\rho_S}\,.
\ee
During slow roll the curvature perturbation of $S$ is slowly changing
with time.

Eliminating $\zeta_\gamma$ and $\rho_\gamma$ we can rewrite
Eqs.~(\ref{curvature1},\ref{curvature2}) in dimensionless form
\bea
\zeta' &=& -\frac{1}{-4+\Omega_\phi +\Omega_S(1-3\omega)} 
\bigg\{ \Big( - \Omega_\phi (3+g/h) 
+ 3 \Omega_S(1+\omega)(3\omega-1) \Big) \zeta \nonumber\\
&& -3 \Omega_S(1+\omega)(3\omega-1) \zeta_S 
+(3+g/h) \Omega_\phi \zeta_\phi \bigg \}\,,
 \\
\zeta_\phi' &=& 
\Big(-4+\Omega_\phi +\Omega_S(1-3\omega) \Big) \frac{-g}{2(g+3h)} \zeta +
\frac{g'}{g+3h} \zeta_S \nonumber \\
 && + \frac{1}{g+3h} \Big( -g'+ \frac{g}{2}(-4+\Omega_\phi + 
\Omega_S(1-3\omega)) \Big) \zeta_\phi \,,
\eea
where as before a prime denotes differentiation w.r.t. the number of
e-foldings $N = \ln a$.


\section{Physical situations}

To simplify and explore the properties of the evolution equations for
the perturbations we will discuss two special cases in this section.

\subsection{$S$ non-dynamical}

First we consider the case where the field $S$ is strictly frozen and
is non-dynamical. In addition we require $S$ to have negligible energy
density.  The evolution of the perturbation is governed by the
fluctuating inflaton decay rate, irrespective of what is the cause for
the fluctuations.  Note that although $\zeta_S$ becomes ill defined in
this limit, the combination $g'\zeta_S = \delta \Gamma$ remains
finite. The decay rate is independent of time. It is then useful to
introduce the quantities
\be
f = \frac{\Gamma}{ H} = \frac{g}{h}\,,
\qquad
\delta_\Gamma \equiv \frac{\delta \Gamma }{\Gamma}\,. 
\ee
In the non-dynamical limit the background equations become:
\bea
\Omega'_\phi &=& \Omega_\phi \left( 1- f - \Omega_\phi\right)\,, \\
f' &=& -\frac{f}{2} (-4+\Omega_\phi)\,.
\eea
The perturbation equations are 
\bea
\zeta' &=& -\frac{(3+f)\Omega_\phi}{-4+\Omega_\phi} 
( \zeta_\phi- \zeta)\,, \\
\zeta_\phi' &=& 
\frac{(-4+\Omega_\phi)f}{2(3+f)} (\zeta_\phi-\zeta) 
-\frac{f \delta_\Gamma}{3+f}\,. 
\eea

\begin{figure}[t]
\centering
\hspace*{-5.5mm}
\leavevmode\epsfysize=8cm \epsfbox{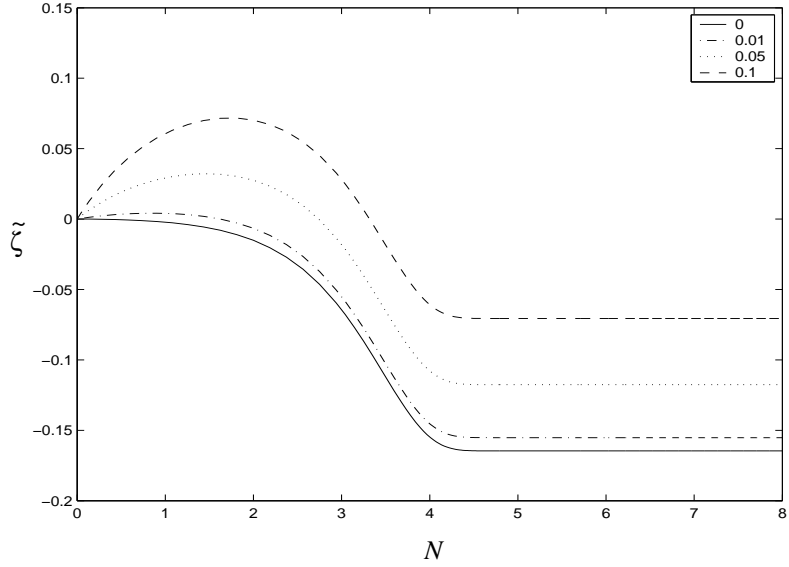}\\
\caption[fig.1] {$\tilde{\zeta}$ as a function of $N = \ln a$ for
different initial inflaton perturbations. The bottom line corresponds
to $\tilde\zeta_{\phi}(0)=0$ (solid). The lines above are for
$\tilde\zeta_{\phi}(0)=0.001$ (dashed-dotted),
$\tilde\zeta_{\phi}(0)=0.05$ (dotted), and $\tilde\zeta_{\phi}(0)=0.1$
(dashed) respectively.}
\end{figure}

We solved the equations numerically.  The results are shown in Fig.~1.
Plotted is the ratio $\tilde{\zeta} \equiv \zeta / \delta_\Gamma$.
This quantity is independent of the value of $\delta_\Gamma$, as
expected~\cite{Dvali}.  The solid line corresponds to negligible
initial curvature perturbation of the inflaton:
$\tilde{\zeta}_\phi(N_0 = 0) = 0$, where we have introduced the
notation $\tilde{\zeta}_\alpha \equiv \zeta_\alpha / \delta_\Gamma$.
In all plots, $f(0) = 10^{-2}$, i.e., the evolution starts at $H =
10^2\,\Gamma$. $H \sim \Gamma$ occurs at $N \approx 3$. We find
$\tilde{\zeta} \approx - 0.166 = - (1 / 6)$.  This agrees with the
results of~\cite{Dvali}, who found for the gravitational potential
during radiation domination $\psi =-(2/3)\zeta
=(1/9)\delta_{\Gamma}$~\footnote
{The metric potentials $\psi$ and $\zeta$ are related to each other by
$\zeta =-(2/3)(H^{-1}\dot\psi+\psi)/(1+\omega)-\psi$. For constant
$\psi$ and $\omega=1/3$, corresponding to radiation domination,
$\zeta=-(3/2)\psi$. For details see Ref.~\cite{Brandenberger,Dvali2}.}
.

Furthermore, we looked at the effects of a non-zero initial
inflaton curvature. For $\tilde{\zeta}_\phi(0) \lesssim 10^{-3}$
the final curvature is indistinguishable from $\tilde{\zeta}_\phi(0)=0$.  
But for larger initial curvatures deviation from 
$\zeta =-\delta_\Gamma /6$ arise. For $\tilde{\zeta}_\phi(0) = 10^{-2}$ the
deviation is $\sim 5\%$, whereas for $\tilde{\zeta}_\phi(0) = 5 \times
10^{-2}$ it is $\sim 30\%$.


\subsection{S oscillating}

It is possible that the inflaton decays while the flat direction field
is oscillating.  This is the case for $\Gamma \lesssim m_S$.  Averaged
over one oscillation $\omega = 0$, and the averaged decay rate is
constant in time. The dimensionless background equations for this case
are
\bea
\Omega'_\phi &=& \Omega_\phi 
\left( 1- \frac{g}{h} - \Omega_\phi - \Omega_S \right)\,, \\
\Omega'_S &=&  \Omega_S \left( 1 - \Omega_\phi - \Omega_S \right) \,,\\
f' &=& - \frac{f}{2} (-4+\Omega_\phi + \Omega_S)\,.
\eea
The perturbation equations are
\bea
\zeta' &=& -\frac{ \left( - \Omega_\phi (3+f) - 3 \Omega_S \right) \zeta 
+ 3 \Omega_S \zeta_S +(3+f) \Omega_\phi \zeta_\phi}{-4+\Omega_\phi +\Omega_S}\,,
\label{osc_zeta}
 \\
\zeta_\phi' &=& 
\frac{(-4+\Omega_\phi +\Omega_S)f} {2(f+3)} (\zeta_\phi-\zeta) 
-\frac{f \delta_\Gamma}{3+f}\,, \\
\zeta'_S &=& 0\,.
\eea
Since $S$ has a fixed equation of state in this case, $\zeta_S$ is
constant, and it is straightforward to incorporate the evolution of
$S$. It is easy to see that when $\Omega_S$ and $\Omega_S~\zeta_S$ are
negligible, the above equations are the same as in the non-dynamical
limit, and the result $\zeta =-\delta_\Gamma /6 $ is obtained.  This
is shown by the solid line in Fig.~2. For $\Omega_S$ large enough, the
final curvature starts to deviate from this value.  A large $\Omega_S$
is possible especially in the scenario discussed in \cite{Dvali2},
where it is not the inflaton but its decay products that have a
varying decay rate.

\begin{figure}[t]
\centering
\hspace*{-5.5mm}
\leavevmode\epsfysize=8cm \epsfbox{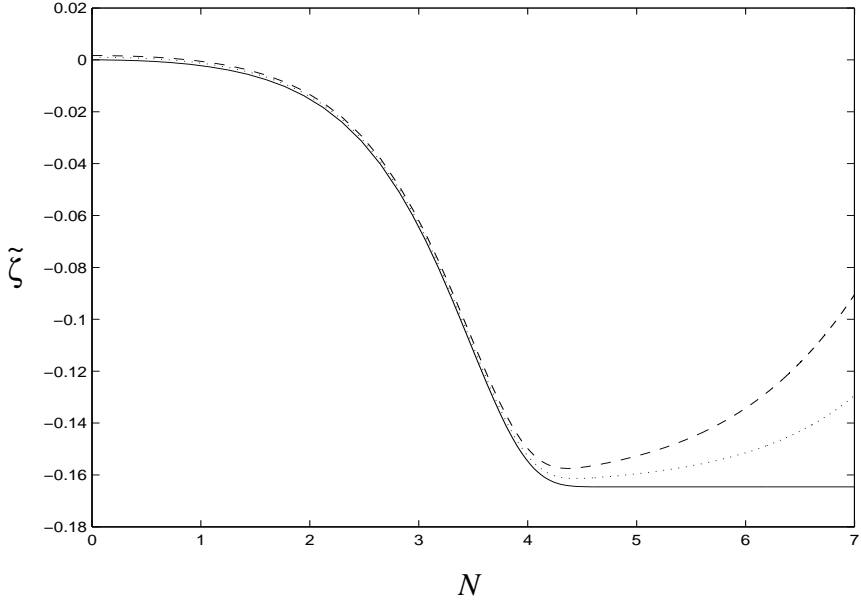}\\
\caption[fig.1] {$\tilde{\zeta}$ as a function of $N = \ln a$ for
different contributions of $\Omega_S$ and $\zeta_S$. The solid line
corresponds to $\Omega_S$ negligible small, the dotted line
corresponds to $\Omega_S (N=0) = 10^{-5}$ and $\tilde{\zeta}_S =
10^{2}$ (for direct decay), and the dashed line corresponds to
$\Omega_S (N=0) = 5 \times 10^{-3}$ and $\tilde{\zeta}_S =1/3$
(indirect decay). In all plots $\delta_\Gamma = 10^{-5}$.}
\end{figure}

The field $S$ does not undergo any transfer of energy: $Q_{S}=0$.  As
a result, a non-negligible energy density $\Omega_S$ can only affect
the total curvature perturbation as a result of non-adiabatic,
relative pressure perturbations, see Eq.~(\ref{rel}).  Since we assume
that the inflaton red shifts as non-relativistic matter after
inflation, there is no relative pressure perturbation between the
inflaton and the $S$ field: $S_{\phi S}=0$. There is a relative
pressure perturbation though between the radiation bath and $S$, and
$S_{\gamma S} \neq 0$.  However, it is only after inflaton decay that
the energy density stored in the radiation bath becomes appreciable,
and that $\Omega_S$ increases. Then, the relative pressure
perturbation leads to the conversion of isocurvature perturbations to
curvature perturbations. This is the idea behind the curvaton
scenario~\cite{curvaton}. The final curvature is approximately given by the
function of both $\zeta_\gamma$ and $\zeta_S$:
\be
\zeta \approx (1 - r) \zeta_\gamma + r \zeta_S
\label{zeta_curv}
\ee
with
\be
r = \frac{3 \Omega_S}{4 \Omega_\gamma + 3 \Omega_S} \,\Bigg|_{S \; {\rm decay}}
\ee
Here $r$ is to be evaluated at the time of $S$-decay. Since $S$ does
not undergo energy exchange it acts like a spectator field during
inflaton decay, and $\zeta_\gamma \approx -\delta_\Gamma /6$,
independently of the value of $\Omega_S$ and decay channels, i.e.
direct/indirect decay modes, see Eq.~(\ref{rate}). 

The curvature perturbation of the $S$-field, $\zeta_S$, can be much
larger for direct decay than for either indirect decay or the
fluctuating mass case, see Eq.~(\ref{decay}). The curvature perturbation
of the flat direction field during the epoch of $S$-oscillations is
\be
\zeta_S = \frac{\delta \rho_S } {3 \rho_S} = \frac{2 \delta S}{3 S}\,.
\ee
For direct decay 
$\delta_\Gamma={\delta \Gamma}/{\Gamma} = {2\delta S}/{M} < \zeta_S$, 
where the last inequality follows from the fact that
the cutoff is larger than the VEV of $S$: $S < M$.  A Gaussian
perturbation spectrum, as indicated by observations, requires 
$\delta S \ll S$, and therefore $\zeta_S \ll 1$.  For indirect decay 
$\delta_\Gamma = 3\zeta_S$, whereas for the fluctuating mass case 
$\delta_\Gamma = 3\zeta_S/2$.

The results for a non-negligible $\Omega_S$ are shown in Fig.~2.  The
solid line corresponds to $\Omega_S$ negligible small, when $r
\zeta_{S}\ll (1-r)\zeta_{\gamma}$, we obtain $\zeta=-\delta_\Gamma /
6$ as before.  The dotted line is for $\tilde{\zeta}_S =10^{2}$ and
$\Omega_S \approx 10^{-5}, \; 2 \times 10^{-5},\; 5\times 10^{-4}$ at
$N = 0,\; 4,\;7$ respectively. Such large curvature perturbations of
$S$ are possible for direct decay. The dashed line shows the plot for
indirect decay, with $\tilde{\zeta}_S= 1/3$ and $\Omega_S \sim 5
\times 10^{-3}, \; 0.1,\;10^{-2}$ at $N = 0, \; 4, \; 7$.  In both
cases the final curvature is well approximated by
Eq.~(\ref{zeta_curv}), with $\zeta_\gamma=-\delta_\Gamma /6$.  The
contribution of $\zeta_S$ to the total curvature rises above the one
percent level for $(\Omega_S \zeta_S)/\zeta_\gamma \gtrsim 10^{-2}$.
For indirect decay and the fluctuating mass case, this is only
possible for $\Omega_S \gtrsim 10^{-2}$.


\section{Conclusions}

In this paper we presented a gauge invariant formalism to study the
evolution of curvature perturbations during decay of the inflaton or
some other field that dominates the energy density.  We specialized to
the case where the perturbations arise from a spatially fluctuating
decay rate.  

The system we considered consists of the inflaton $\phi$, the
radiation bath, and a light fluctuating field $S$ which is responsible
for the fluctuations in the decay rate of $\phi$.  During inflaton
decay there is a energy transfer from the inflaton field to the
radiation bath.  Due to the fluctuating decay rate this energy
transfer has an intrinsic non-adiabatic component, which feeds the
curvature perturbation of the radiation bath. This can be understood
intuitively from the fact that fluctuations in the inflaton decay rate
leads to fluctuations in the reheat temperature of the Universe, and
therefore to energy fluctuations in the thermal bath $\zeta_\gamma
\propto \delta \rho_\gamma /\rho_\gamma \propto \delta \Gamma /
\Gamma$.  Since after inflaton decay the universe is radiation
dominated, the total curvature perturbation is 
$\zeta \approx \zeta_\gamma$.

We studied the evolution of the background fields and the
perturbations numerically. We found that the total curvature
perturbation is given by $\zeta=-(1/6)\delta_{\Gamma}$ and $\psi
=(1/9)\delta_{\Gamma}$. Here we have assumed that initially
$\zeta_\phi$ and $\Omega_S$ are negligible small.  Departure from
$\zeta=-(1/6)\delta_{\Gamma}$, above the one percent level, arises for
larger values of $\zeta_{\phi}(0)/\zeta_\gamma \gtrsim 10^{-2}$ and/or
$\Omega_S \zeta_S / \zeta_\gamma \gtrsim 10^{-2}$.  It is interesting
to note that for indirect decay $\zeta_S$ can be large: $\zeta_S \gg
\delta_\Gamma$ if the VEV of the flat direction field is much less
than the cutoff scale $S \ll M$.

\section{Note}
While we were finishing our work we noticed a related contribution by
Matarrese and Riotto, see \cite{Matarrese}. In this paper, the gauge
invariant curvature perturbations for $\phi, \gamma$ are
presented. The authors have analytically obtained
$\psi=(4/45)\delta\Gamma/\Gamma$, which in our case is proven
numerically to be $\psi =(1/9)\delta\Gamma/\Gamma$.

\section{Acknowledgments}
The authors are thankful to Robert Brandenberger, Kari Enqvist, and 
David Wands for helpful discussions. M. P. is supported by the European
Union under the RTN contract HPRN-CT-2000-00152 Supersymmetry in the
Early Universe, and A. M. is a CITA-national fellow.

 


\begin{thebibliography}{99}

\bibitem{Linde}
For a review, see: A.D. Linde, {\it Particle Physics And Inflationary
Cosmology}, Harwood (1990).

\bibitem{curvaton}
S. Mollerach, Phys. Rev. D {\bf 42}, 2 (1990).
K.~Enqvist and M.~S.~Sloth,
Nucl.\ Phys.\ B {\bf 626}, 395 (2002)
[arXiv:hep-ph/0109214].
T.~Moroi and T.~Takahashi,
Phys.\ Lett.\ B {\bf 522}, 215 (2001)
[Erratum-ibid.\ B {\bf 539}, 303 (2002)]
[arXiv:hep-ph/0110096].
D.~H.~Lyth and D.~Wands,
Phys.\ Lett.\ B {\bf 524}, 5 (2002),
[arXiv:hep-ph/0110002];




\bibitem{Dvali}
G. Dvali, A. Gruzinov, and M. Zaldarriaga, astro-ph/0303591.

\bibitem{Kofman}
L. Kofman, astro-ph/0303614.

\bibitem{epm}
K.~Enqvist, A.~Mazumdar and M.~Postma,
arXiv:astro-ph/0304187.

\bibitem{Dvali2}
G.~Dvali, A.~Gruzinov and M.~Zaldarriaga,
arXiv:astro-ph/0305548.


\bibitem{Mazumdar}
A.~Mazumdar,
arXiv:hep-ph/0306026.


\bibitem{Bardeen}
J. M. Bardeen, Phys. Rev. D {\bf 22}, 1882 (1980);
J. M. Bardeen, P. J. Steinhardt and M. S. Turner, Phys. Rev. D {\bf 28}, 679
(1983);
J. M. Bardeen, DOE/ER/00423-01-C8 {\it Lectures given at 2nd Guo-jing Summer School on
particle physics and cosmology, Nanjing, China, Jul 1988}.

\bibitem{Brandenberger}
V.~F.~Mukhanov, H.~A.~Feldman and R.~H.~Brandenberger,
Phys.\ Rept.\  {\bf 215}, 203 (1992).


\bibitem{Bellido}
J.~Garcia-Bellido and D.~Wands,
Phys.\ Rev.\ D {\bf 53}, 5437 (1996)
[arXiv:astro-ph/9511029].



\bibitem{Kodama}
H. Kodama, M. Sasaki, Prog. Theor. Phys. Suppl. {\bf 78}, 1 (1984).

\bibitem{Wands2}
K.~A.~Malik, D.~Wands and C.~Ungarelli,
Phys.\ Rev.\ D {\bf 67}, 063516 (2003)
[arXiv:astro-ph/0211602].

\bibitem{Matarrese}
S. Matarrese and A. Riotto, astro-ph/0306416.






\end{thebibliography}
\end{document}